\documentclass[journal]{IEEEtran}
\IEEEoverridecommandlockouts
\usepackage{enumerate,url,diagbox,amsmath,amsthm,amssymb,graphicx,color,cite,algorithm,algpseudocode,algorithmicx}

\floatname{algorithm}{Algorithm}

\begin{document}
\title{Meta-Key: A Secure Data-Sharing Protocol under Blockchain-Based Decentralised Storage Architecture}
\author{
\IEEEauthorblockN{Dagang Li\IEEEauthorrefmark{1}\IEEEauthorrefmark{2}, Rong Du\IEEEauthorrefmark{1}, Yue Fu\IEEEauthorrefmark{1}, Man Ho Au\IEEEauthorrefmark{3}}\\
\IEEEauthorblockA{\IEEEauthorrefmark{1}School of Electronic and Computer Engineering, Peking University, China}\\
\IEEEauthorblockA{\IEEEauthorrefmark{2}PKU-HKUST ShenZhen-HongKong Institution, Shenzhen, China}\\
\IEEEauthorblockA{\IEEEauthorrefmark{3}Department of Computing, The Hong Kong Polytechnic University, Hong Kong, China}\\
Email: fuyuefyu@126.com
\thanks{Corresponding author: Yue Fu. This work was supported by Shenzhen Key Lab Project (ZDSYS201703031405137) and the Shenzhen Municipal Development and Reform Commission (Disciplinary Development Program for Data Science and Intelligent Computing)}
}
\maketitle
\begin{abstract}In this letter we propose Meta-key, a data-sharing mechanism that enables users share their encrypted data under a blockchain-based decentralized storage architecture. All the data-encryption keys are encrypted by the owner's public key and put onto the blockchain for safe and secure storage and easy key-management. Encrypted data are stored in dedicated storage nodes and proxy re-encryption mechanism is used to ensure secure data-sharing in the untrusted environment. Security analysis of our model shows that the proxy re-encryption adopted in our system is naturally free from collusion-attack due to the specific architecture of Meta-key.
\end{abstract}
\section{Introduction}
Security and reliability of traditional centralized cloud-storage architecture rely solely on the cloud-service provider. Data hold by third-party servers may be eavesdropped, stolen or destroyed by politic, technological or legal means. Users are unable to get their data when vendors stop their service.\\
\indent A blockchain-based decentralized storage system can offer service without dependence on a specific vendor: nodes in the network contribute their disk space to store data for others, each node can be space demander, provider or both. Data are encrypted and cipher-texts are distributed to anonymous nodes in the network so security is also strengthened by the hiding of location. The location information is encrypted and put to a blockchain maintained by all nodes as meta-data. Such mechanism realizes a true decentralized storage: data are managed and stored among all nodes securely, from whom owners can access their data any time they want.\\
\indent However, data-sharing becomes rather cumbersome under such architecture for two reasons: first, traditional data-sharing mechanisms are not applicable to this architecture; second, sharing of encrypted data requires compatible key management protocol for blockchain. Aimed at the two problems we propose Meta-key as a feasible key-management and data-sharing mechanism compatible with blockchain-based decentralized storage. Proxy re-encryption is introduced to realize cipher-text transformation and restoration to solve the security issues of key-sharing under untrustworthy environments. A security model is constructed to prove the collusion-free property of Meta-key.

\indent The contributions of this letter are summarized as the following:
\begin{enumerate}
\item \textbf{Meta-key.} We propose a Meta-key mechanism, where data decryption keys are stored in a blockchain as part of the metadata and protected by user's private key. This efficiently realizes an easy and secure key-management mechanism in a decentralized fashion.
\item \textbf{Secure proxy re-encryption.} We prove that Meta-key is naturally free from collusion-attack under untrusted environments, even if the adopted proxy re-encryption scheme for secure data-sharing doesn't hold this property.
\end{enumerate}

\indent The rest of the letter is organized as following. In section \uppercase\expandafter{\romannumeral2}, some background and related work are introduced briefly. In section \uppercase\expandafter{\romannumeral3}, details of Meta-key mechanism will be elaborated. In section \uppercase\expandafter{\romannumeral4}, security discussion and proof of our protocol will be given. Finally, the conclusions will be drawn in section \uppercase\expandafter{\romannumeral5}.
\section{Background \& Related work}Blockchain technology\cite{1} can be directly applied to cloud-storage architecture. Due to the large volume of the data, only meta-data of them can be stored on-chain, protected by the owner's private key. When data are stored, data-owner chooses a feasible location among the nodes and put his encrypted data there, and the location information is put on-chain as part of the meta-data. When the owner wants to read back his data, he can retrieve the meta-data from the blockchain and decrypt with his private key to reveal the data-location. Then, he downloads the data from the corresponding nodes.\\
\indent Nowadays, many blockchain-based cloud-storage systems are coming to the fore, such as Storj\cite{2}, Enigma\cite{3}, Metadisk\cite{4}. Xunlei Network Corp. also encourages its users to share their idle bandwidth \& hard disk resources by Onethingcloud (a small device linked to family routers) and reward them with Linktoken\footnote{https://www.onethingcloud.com/uk/site/index.html}. As of this writing, Xunlei claimed that over 1,500,000 nodes share 1500PB of storage for them voluntarily\footnote{http://www.onethingtech.net/news/20180522.php}.

\section{Meta-key mechanism}Data-sharing turns out to be a problem under blockchain-based storage architecture. Existing data-sharing strategies are based on centralized cloud-service providers. Data-owners simply authorize the trusted service providers and let them carry out the sharing process. However, in blockchain storage where every node is untrustworthy, existing data-sharing schemes can not be applied directly. We need a secure data-sharing scheme that can work under blockchain-based storage architecture.\\
\indent Furthermore, if data are shared in their encrypted form, owners must provide their decryption key to others, which either undermines the security of the owner's other data in an untrusted environment or we need a better key-management mechanism to encrypt each piece of data with a separate key. In this section, we propose a series of policies as an attempt to solve these problems. The definition of symbols used in this letter is given in Table \uppercase\expandafter{\romannumeral1}.
\begin{table}[!h]
\newcommand{\tabincell}[2]{\begin{tabular}{@{}#1@{}}#2\end{tabular}}
\centering
\caption{Symbol Definition}
\begin{tabular}{cc}
\hline
Symbol & Definition\\
\hline
M & Meta-data\\
C & Original data-ciphertext\\
C' & Re-encrypted data-ciphertext\\
S & Original encryption key kept by data-owner A\\
S' & \tabincell{c}{Generated decryption key for recipient B \\to decrypt re-encrypted ciphertext}\\
R & Proxy re-encryption key, generated from S and S'\\
$N_{1}$ & Location of original data-ciphertext\\
$N_{2}$ & Location of randomly picked data-sharing node\\
PRec & Proxy Re-encryption\\
Gen & Key generation\\
Dec & Decryption\\
\hline
\end{tabular}
\end{table}
\subsection{Blockchain as a metadata store}Blockchain is naturally a decentralized storage system maintained by all nodes in the network, which leads to the blockchain bloat problem: every node must store a copy of every transaction in the blockchain so it will soon expand to an unmanageable size when storing large data. As mentioned before, to address this problem, some meta-data of the data are extracted and stored in blocks on-chain rather than the complete data themselves, which may include date, hashing outputs, storage location, etc. The full data are stored into dedicated storage nodes off-chain. Both data and meta-data are encrypted by the data-owners\cite{4}.\\
\indent Secure data-sharing of a general cloud storage system is commonly performed in the following way. When any recipient sends a sharing-request, the data-owner encrypts his data and uploads the data cipher-text to cloud-storage. Then, he applies that recipient's public key to encrypt the decryption key and sends the key cipher-text to the recipient through some secure channels. The recipient decrypts the key cipher-text with his private key to get the decryption key and then download the data cipher-text to decrypt.\\
\indent In the untrusted environment we have, the only secure and reliable channel for decryption key transport will be the blockchain, therefore in Meta-key we put the key cipher-text also on-chain as a part of the meta-data, and use the blockchain for both key-management and key-distribution, and thus comes the name of our proposal. Since the meta-data on-chain are protected by the owner's private key, a secure data-sharing now turns into the generation of a new record of meta-data on-chain owned by the recipient for his copy of the shared data. Fig.1 shows how Meta-key mechanism works compatibly with blockchain-based cloud-storage architecture.

\begin{figure}[!h]
  \centering
  \includegraphics[width=3.0in]{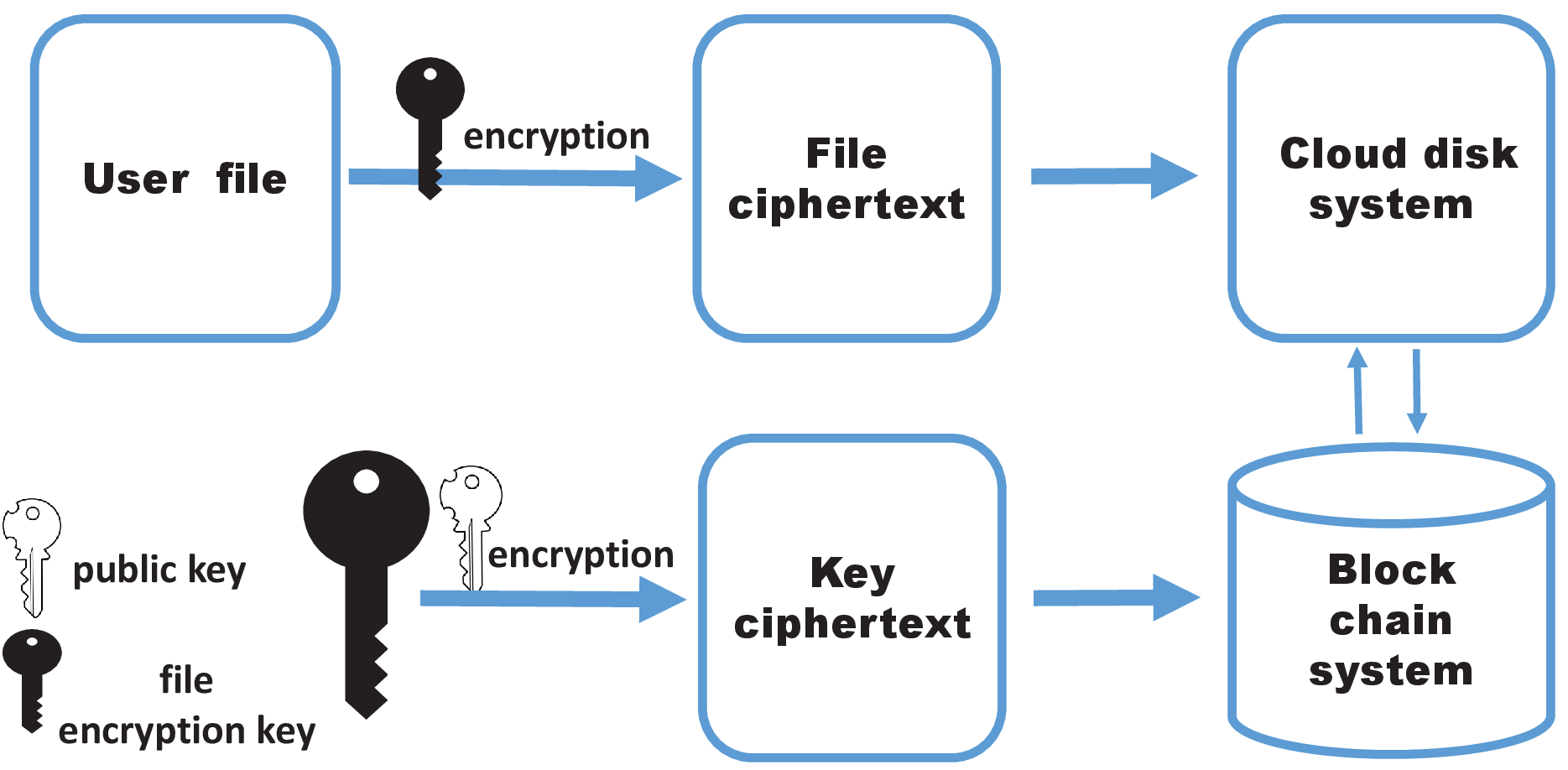}
  \caption{Meta-key mechanism for blockchain-based cloud-storage architecture}
\end{figure}

\indent In a blockchain-based cloud storage, the security of data is both protected by encryption and the conceal of the location. These locations are recorded into meta-data that can only be read by data-owners, therefore we need to use new decryption key and location for the shared data different from the original copy and only known to the recipient to ensure the level of security for both parties after the sharing. Since the data-to-be-shared are very likely already encrypted and stored in the cloud, asking the data-owner to download, decrypt, re-encrypt (with a new decryption key) and upload again (to a different location) is not only very inefficient, but also poses security risks because the plain-text needs to be recovered during the sharing process.\\
\indent To tackle these issues, proxy re-encryption is chosen as the foundation of key and cipher-text transformation mechanism. It will be introduced in the following part.

\subsection{Proxy re-encryption}Proxy re-encryption is a cipher transformation scenario that is widely used in the context of data-sharing in cloud-environment. It was first proposed by Blaze et al. in 1998\cite{5}. Without revealing any information about key or plain-text, it allows a semi-trusted proxy to transfer Alice's cipher-text to Bob's cipher-text with the same plain-text. ``Semi trusted" means the proxy will strictly execute the encryption steps as the algorithm. Ateniese et al. formalized it into strict definition and proposed a series of proxy re-encryption schemes. Application in distributed storage systems are also discussed. It is widely used in many fields such as mail filter\cite{6}, distributed file system management\cite{7} and intellectual property protection\cite{8}.\\
\indent As shown in Fig.2, in traditional cloud-service, it can be applied in this way: suppose Alice and Bob are two users of the same cloud-service provider. Alice uploads her data encrypted by her public key $P_{a}$. Hence, the provider knows nothing about the plain-text. When Alice requests to share her data with Bob, she combines her private key and Bob's public key $P_{b}$ to generate a transformation key $R_{k}$ and sends it to the cloud-service provider. Acting as a proxy, the provider operates the re-encryption with $R_{k}$. Hence, it's easy for Bob to download the re-encrypted cipher-text on-cloud and decrypt it by his private key.

\begin{figure}[h]
  \centering
  \includegraphics[width=3.5in]{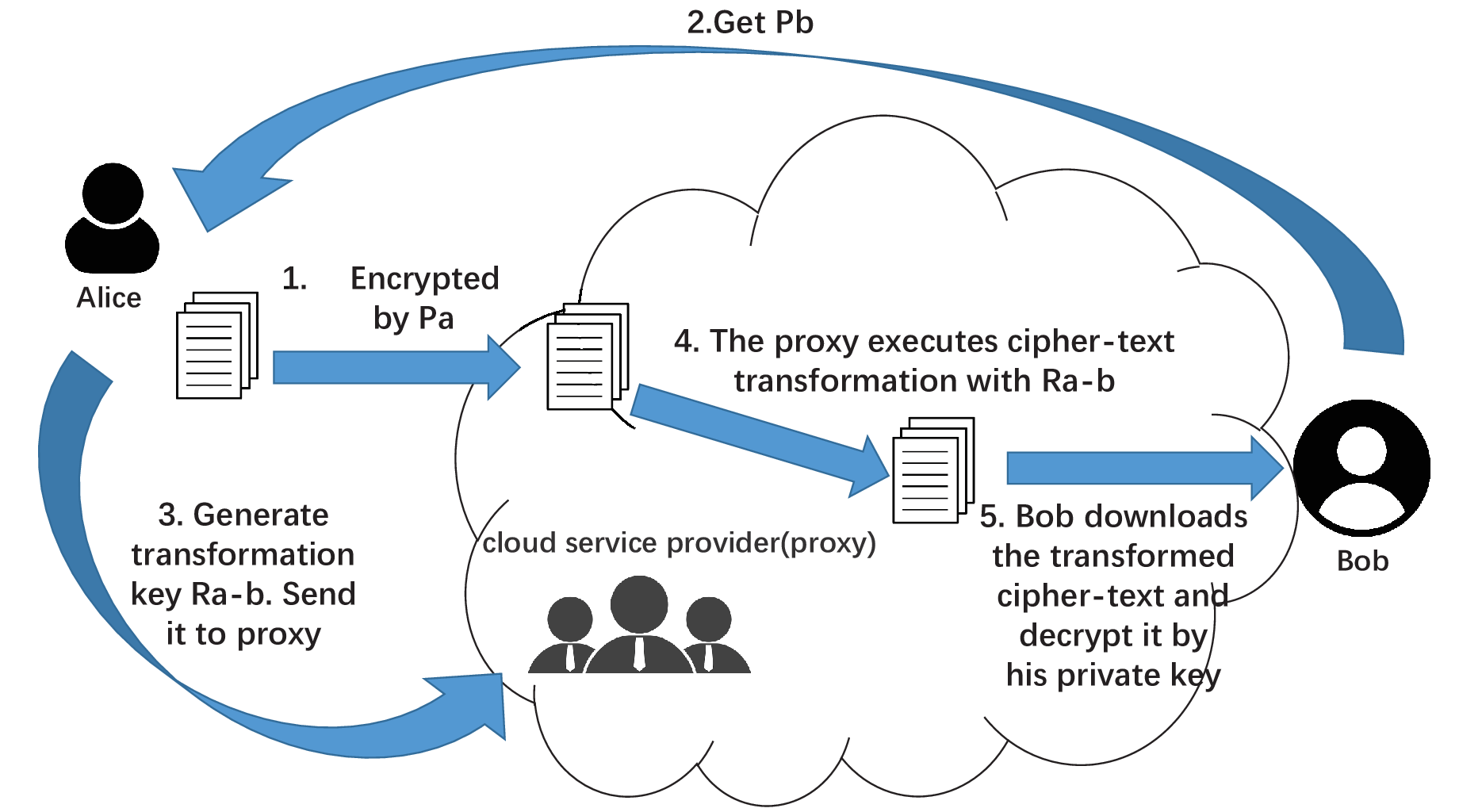}
  \caption{Proxy re-encryption on-cloud.}
\end{figure}

\subsection{Overall process of data-sharing}Our design follows the concept of proxy re-encryption with some modification described below, because here in a decentralized environment we can't completely trust the proxy who might collude with the recipient to attack on the data-owner's private key. We choose to avoid direct interact and keep anonymity between the proxy and the recipient during the whole process of data sharing. 
Furthermore, since in our design different data are encrypted with different keys, we can let Alice choose a random key for the shared data and don't need to bother asking Bob. The new key can be safely given to Bob as part of the on-chain meta-data of Bob's copy. \\
\indent Therefore let $N_{1}$ be the proxy and the data-owner be Alice whose encryption key is S. Alice chooses a new S' and picks a new server $N_{2}$ for the copy of Bob. She generates R from S and S', and put S' and $N_{2}$ on-chain encrypted by Bob's public key. R and $N_{2}$ are sent to the proxy who re-encrypt the cipher-text and store it to $N_{2}$. Bob will get his copy from $N_{2}$ using the meta-data from the blockchain without knowing anything about $N_{1}$.
The detailed data-sharing process is summarized into Algorithm 1. The overall system framework is shown in Fig.3.

\begin{algorithm}[t]
\caption{Data-sharing in Meta-key}
\begin{algorithmic}[1]
\State $Dec(M)\rightarrow S, N_{1}$;
\State Gen S';
\State Gen(S,S')$\rightarrow$ R;
\State Alice send R $\rightarrow N_{1}$;
\State $N_{1}$ PRec (C) $\rightarrow$ C' with R;
\State $N_{1}$ send C' $\rightarrow$ $N_{2}$;
\State Alice send S' \& $N_{2}\rightarrow$ Bob via blockchain;
\State Bob download C' from $N_{2}$, Dec($N_{2}$) with S';
\end{algorithmic}
\end{algorithm}


\begin{figure}[t]
  \centering
  \includegraphics[width=3.5in]{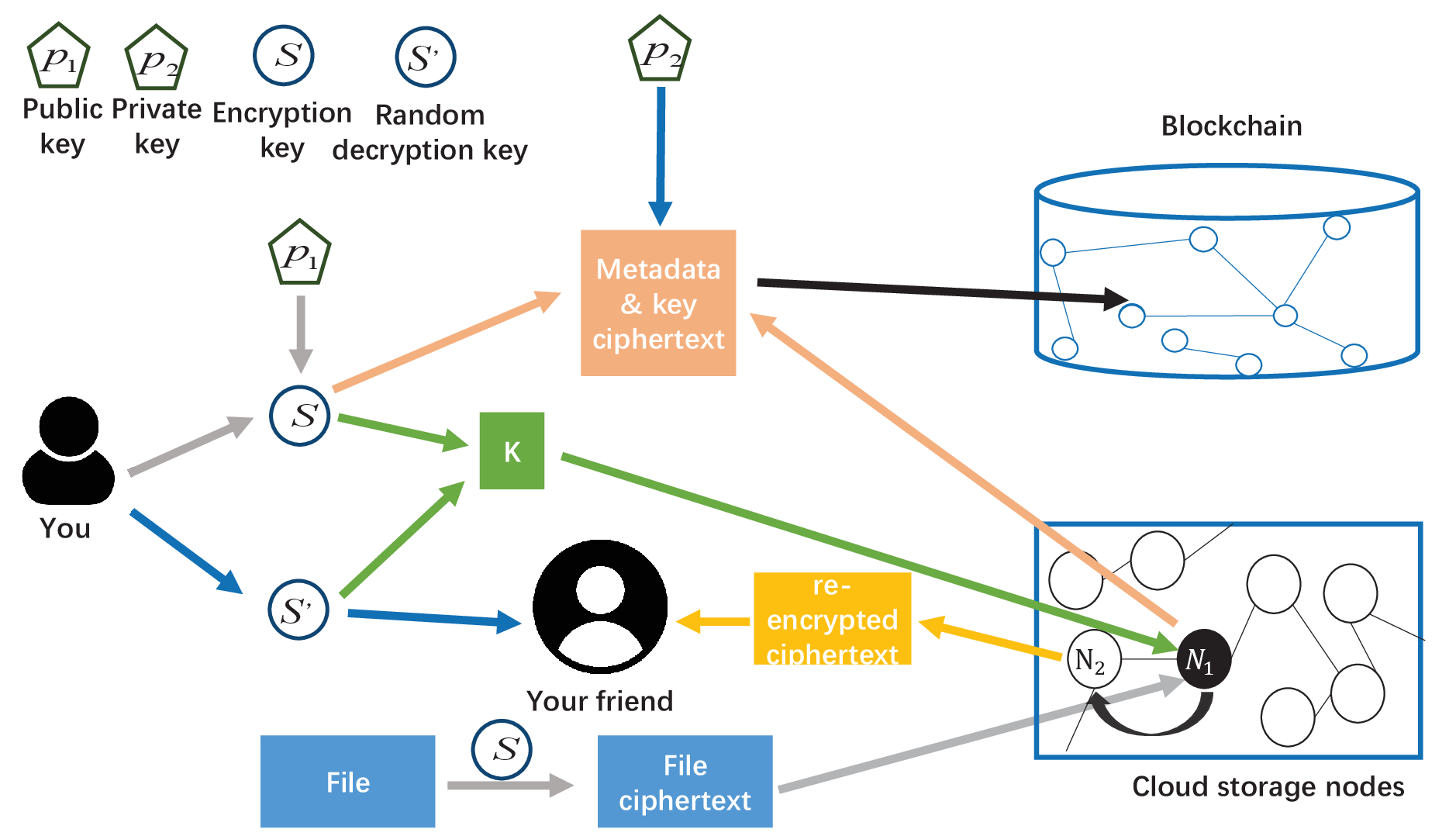}
  \caption{The final data-sharing process.}
\end{figure}

\section{Security analysis}In this section we will analyze the security of Meta-key. We will prove that Meta-key is naturally free from collusion-attack benefiting from its architecture, even if the specific proxy re-encryption scenario does not have such property. 
\subsection{Security model of Meta-key}There are two layers of security in the context of this architecture. The first layer is the indistinguishably of cipher-text location, that is, attackers are not able to determine the owner of a given cipher-text and therefore are not able to distinguish his target cipher-text from others. The second layer is that, even if an attacker can determine where his target cipher-text is and succeed in stealing it, he is still not able to read it without the decryption key. Clearly, layer 2 is general so we will only discuss the security of layer 1 in details. We will start by a series of definitions.

\textbf{Definition 1. Cipher-text location security game.} 
\begin{enumerate}
\item Preparing process. A challenger $\mathcal{C}$ chooses a location $l$ from $n$ possible nodes. He then hides his cipher-text C into $l$. 
\item Assume that there is an attacker $\mathcal{A}$ who is not able to read the ciphertext. He guesses a location $l'$ for his target cipher-text based on the information he owns. 
\item If $l=l'$, we say $\mathcal{A}$ wins this game. The superiority of $\mathcal{A}$ winning this game is further defined as $Pro(l=l')-\frac{1}{n}$, where $Pro$ refers to the probability of an event.
\end{enumerate}
 
\indent With the concept of attacker's superiority, the indistinguishably of cipher-text location security can be defined as follow:

\textbf{Definition 2. Cipher-text location secure (CLS).} A strategy is defined to be cipher-text location secure, if and only if the superiority of any potential attacker $\mathcal{A}$ is ignorable.

\indent In Meta-key, the delegator Alice requests $N_{1}$, the node where the original cipher-text C is stored, to re-encrypt C to C' as a proxy and to send C' to $N_{2}$, then the delegatee Bob can read C' from $N_{2}$. Clearly, in this process the location of $N_{2}$ must be revealed to both Bob and $N_{1}$ by Alice, therefore neither C for $N_{1}$ nor C' for Bob and $N_{1}$ is CLS. Then we have the lemma:

\textbf{Lemma 1.} C for Bob and $N_2$, C' for $N_2$ are both CLS.\\
\indent \textbf{Proof.} It's clear that both Bob and $N_2$ has no information about the location of C so they are CLS. For $N_2$, even though it knows C' comes from $N_1$, he doesn't know it is actually from Alice. Besides, C and C' are indistinguishable to $N_2$ if it can not read the cipher-text. Hence, C' is also CLS to $N_2$.$\hfill\blacksquare$
\subsection{The collusion attack}So far the CLS property of Meta-key is defined and discussed. The existing CLS of C and C' for a single node is also shown in Lemma 1. However, nodes may conspire trying to gain more information of locations and identities. Furthermore, in the context of proxy re-encryption, a collusion attack may be constructed between malicious nodes: Alice's decryption key S can be calculated with the knowledge of R, C, C' and Bob's decryption key S'. Hence, when the proxy who knows R, C colludes with Bob who knows C' and S', S is at risk of being revealed.

\textbf{Definition 3.} A pair of nodes is called motivated co-conspirators, if and only if the benefit increased on superiority is greater than the potential cost paid to reach a collusion, where the potential cost refers to the level of difficulty to reach each other out of anonymity.\\
\indent With definition 3, we have the following theorem:

\textbf{Theorem 1.} For delegatee Bob, proxy $N_1$, data-sharing node $N_2$, who may be malicious, they are not motivated co-conspirators to each other in the context of CLS.\\
\indent \textbf{Proof.} Consider the collusion between Bob+$N_2$ and $N_1+N_2$. According to Lemma 1, both C and C' are CLS to $N_2$ so $N_2$ reveals no information to its conspirators. Hence, combination of Bob+$N_2$ and $N_1+N_2$ are not motivated co-conspirators.\\
\indent For $N_1$+Bob, they can reveal information of C and C' to each other. However, $N_1$ only known to Alice, whose location is totally random to Bob, vice versa. Suppose it makes no difference to request and verify the following two questions: ``Are you $N_1$ holding C, chosen by Alice?", ``Are you the delegatee Bob chosen by Alice?" The difficulty of their collusion equals to an attacker of a CLS game with no superiority. Hence, they are not motivated co-conspirators.$\hfill\blacksquare$\\
\indent Now let us discuss the collusion attack in the context of proxy re-encryption. In Meta-key, the role of proxy is in fact divided to $N_{1}$ and $N_{2}$. $N_{1}$ re-encrypts C by R and sends C' to $N_{2}$ and Bob gets C' from $N_2$. Hence, $N_1$ holds C,C' and R, whereas Bob holds S'. Besides, the location of $N_2$ should be revealed to $N_1$ and Bob by Alice. The knowledge of nodes are induced in Table \uppercase\expandafter{\romannumeral2}.

\begin{table}[!t]
\newcommand{\tabincell}[2]{\begin{tabular}{@{}#1@{}}#2\end{tabular}}
\centering
\caption{Knowledge of nodes, including location of nodes}
\begin{tabular}{cc}
\hline
Node & Information\\
\hline
Bob & S', $N_{2}$\\
$N_{1}$ & R, C, C', $N_{2}$\\
$N_{2}$ & C'\\

\hline
\end{tabular}
\end{table}

\indent From Table \uppercase\expandafter{\romannumeral2} we can see, similar to theorem 1, Bob and $N_1$ are not motivated to conspire with $N_2$: though location of $N_2$ is revealed to Bob and $N_1$, $N_2$ is able to provide neither C for Bob, nor S' for $N_1$. Therefore, only collusion between $N_1$ and Bob should be discussed: $N_1$ doesn't know anything about S' held by Bob, nor does Bob know anything about C held by $N_1$. The collusion will not happen if these information are not exchanged. Hence, the hardness for such collusion is at least equals to the level of difficulty to reach each other out of anonymity. We have the following theorem:

\indent\textbf{Theorem 2.} Even if the proxy re-encryption strategy is not collusion-free secure, the hardness for collusion between proxy and delegatee is at least no less than a CLS game.
\indent \textbf{Proof.} It is a natural deduction of theorem 1. $\hfill\blacksquare$\\

\indent All in all, if the brute-force solution for a CLS game deems to be hard, we've proven that Meta-key is naturally free from collusion attack, benefiting from its architecture.

\subsection{Reliability of data}The availability and reliability of on-chain meta-data is guaranteed by the blockchain. However, reliability of off-chain data cipher-text may still be at risk. Though encrypted, they may still be distorted or lost in untrusted $N_{1}$s hence they must be redundantly stored. In Metadisk\cite{4} simple replications are adopted, where copies of $C'$ are sent to several $N_{1}$s. Hash authentication is applied to ensure the completeness of C'. When any replication is corrupted, the failed node requests other surviving nodes for repairing.\\
\indent Erasure codes can be further introduced to enhance the security and reliability of data cipher-text, where ciphers are re-encoded, split into pieces and redundantly stored in various of nodes. The encoded data-shares can still be transferred, proxy re-encrypted and repaired as we've described in the Meta-key model. We just need to recombine enough data-shares collected from surviving nodes. However, detailed discussions are beyond the scope of this letter.

\section{Conclusion}In this letter, we proposed a Meta-key based approach for secure data sharing in a decentralized storage system based on blockchain. We focused on the collusion-free property of the proposed cryptographic protocol and proved it strictly.

\section*{Acknowledgement}Special thanks to Prof. Chunming Tang in Guangzhou University, who generously provided constructive discussions on proxy cryptography.

\end{document}